\documentstyle[preprint,aps]{revtex}
\begin{document}
\author{Angel S\'anchez}
\address{
Escuela Polit\'ecnica Superior,
Universidad Carlos III,
Avda.\ Mediterr\'aneo 20, 28913 Legan\'es, Madrid,
Spain}
\author{Francisco Dom\'\i nguez-Adame}
\address{
Departamento de F\'\i sica de Materiales,
Universidad Complutense,
Ciudad Universitaria,
28040 Madrid, Spain}
\date{\today}
\maketitle
\begin{abstract}
We consider a one-dimensional continuous (Kronig-Penney) extension of
the (tight-binding) Random Dimer model of Dunlap {\em et al.}\ [Phys.\
Rev.\ Lett.\ {\bf 65}, 88 (1990)]. We predict that the continuous model
has infinitely many resonances (zeroes of the reflection coefficient)
giving rise to extended states instead of the one resonance arising in
the discrete version. We present exact, transfer-matrix numerical
calculations supporting, both realizationwise and on the average, the
conclusion that the model has a very large number of extended states.
\end{abstract}
\narrowtext

The question as to whether Anderson localization \cite{Anderson} occurs
for any one-dimensional, disordered system, has been considered solved
for many years. However, a number of recent papers strongly suggest that
the widely held view that disorder prevents long-range transport in
one-dimension is not always true. This is the case with the work by
Dunlap,
Wu, and Phillips \cite{Dunlap,Wu1,Wu2} (see \cite{Phillips} for a
review). These authors studied a tight-binding model (so called Random
Dimer model, RDM) in which each on-site energy can have only one of two
possible
values, one of these values being assigned at random to {\em pairs} of
lattice sites. They showed that for a certain energy the reflection
coefficient of a {\em single} defect vanished, and that this resonance
was preserved when a finite concentration of defects were randomly
placed in the chain. This
gave rise to a set of delocalized states proportional to the square root
of the number of sites. As a consequence, in such a system electronic
transport can take place almost balistically. Similar results have been
also shown to hold true for dilute binary alloys by Flores \cite{Flores}.
Besides, the arguments of Dunlap {\em et al.}\ to conclude that such a
large number of states were not localized have been further confirmed by
perturbative calculations by Bovier \cite{Bovier}. The RDM has been
generalized recently by Wu, Goff, and Phillips \cite{Wu3} to include
more complex arrangements of defects. In this work, they also considered
the continuum limit of their model, given by square barriers and wells
randomly placed on a line, and it appeared to exhibit the same kind of
localization-delocalization transition.

In this letter, we concern ourselves with an even simpler continuous
model, which we call continuous Random Dimer model (henceafter, CRDM).
This system is the natural continuum version of the RDM of Dunlap {\em
et al.}, although it is not its continuum limit as studied in
\cite{Wu3}. We build
our model in the following way: We start from a usual Kronnig-Penney
model, given by a potential of the form
\begin{equation}
\label{KP}
V(x)=\sum_{n=1}^N\>\lambda_n\>\delta(x-x_n).
\end{equation}
We choose $\lambda_n>0$; the extension of our computations below to the
$\lambda_n<0$ case is straightforward although a bit involved.
To mimic the RDM, we now take the positions of the delta functions to be
regularly spaced, i.e., $x_n=na$, $a$ being the lattice spacing;
moreover, we allow only two values for $\lambda_n$, $\lambda$ and
$\lambda'$, with the additional constraint that $\lambda'$ appears only
in pairs of neighbouring deltas. In this fashion, we have defined what
comes naturally as a continuous version of the tight-binding Random
Dimer model. The corresponding Schr\"odinger equation is then
\begin{equation}
\label{Sch}
\left[-{d^2\mbox{\ }\over
dx^2} + \sum_{n=1}^N\>\lambda_n\>\delta(x-x_n)\right] \Psi (x) =
E\>\Psi(x).
\end{equation}
We believe that Eq.\ (\ref{Sch}) represents a more {\em realistic}
model than the RDM since no tight-binding approach is involved. In
addition, it is well known that the $\delta$-function potential is a
good candidate to model more structured and more sophisticated
interaction potentials \cite{Francisco}. We will see in the following
that there exists a number of energies for which the reflection
coefficient at a single dimer vanishes. Because interference effects are
quite more complex in a continuous model than in a tight binding
approach, it is a nontrivial task to elucidate whether these resonances
will survive when several dimers are located at random along the
lattice.

We can now use the techniques of dynamical systems theory, as first used
by Bellissard {\em et al.}\ \cite{Bellissard} (see also
\cite{Sokoloff}) to construct the Poincar\'e map associated with Eq.\
(\ref{Sch}). It is important to stress that, by doing this reduction
to an equivalent tight-binding set of equations, we are not losing any
information at all, and the calculations remain exact. The resulting
equations are
\begin{equation}
\label{Poincare}
\Psi_{n+1}+\Psi_{n-1} =\left[ 2\cos\sqrt{E}+\lambda_n {\sin\sqrt{E}
\over \sqrt{E}} \right] \Psi_n,
\end{equation}
where $\Psi_n\equiv\Psi(x=na)$. Notice that the energy enters in the
equations in a rather complicated fashion.
To proceed, we have to take into account in the
first place the condition for an electron to be able to move in the
perfect ($\lambda'=\lambda$) lattice, namely
\begin{equation}
\label{constraint}
\left| \cos\sqrt{E}+{\lambda\over 2}\> {\sin\sqrt{E}
\over \sqrt{E}} \right| \leq 1;
\end{equation}
this constraint gives the allowed energy values once $\lambda$ is fixed.
On the other hand, we follow Dunlap {\em et al.} and study the
problem of a single pair defect on an otherwise perfect chain. In our
case, a straightforward application of the results in
\cite{Dunlap} leads to the following condition for
the vanishing of the reflection coefficient:
\begin{equation}
\label{ref}
\cos\sqrt{E}+{\lambda'\over 2}\> {\sin\sqrt{E}
\over \sqrt{E}} =0.
\end{equation}
It is a matter of simple algebra to transform the two equations
(\ref{constraint}) and (\ref{ref}) into these other, more useful
two:
\begin{mathletters}
\label{todas}
\begin{eqnarray}
\label{eq1}
-{2\over\lambda'} & = & {\tan\sqrt{E}\over \sqrt{E}}, \\
\label{eq2}
|\cos\sqrt{E}| & \leq & {\lambda\over|\lambda-\lambda'|}.
\end{eqnarray}
\end{mathletters}
Restricting ourselves to the range $0\leq\lambda'\leq 2\lambda$, Eq.\
(\ref{eq2}) is trivially verified, and therefore it poses no
restrictions on the allowed energy values, aside from the fact that they
must be positive. Hence, we are left only with Eq.\
(\ref{eq1}) to select the energy values for which the
reflection coefficient of a single defect becomes exactly zero.
As the $\tan\sqrt{E}$ is a $\pi$-periodic function and it takes all
values in $[-\infty,+\infty]$, for any $\lambda'$ we choose we
will find energies solving (\ref{eq2}) in {\em each} interval
$[n\>\pi/2,(n+1)\>\pi/2]$, i.e., we will have an {\em infinite countable
set of energies} for which the single defect reflection coefficient
vanishes. This is to be compared with the result of Dunlap {\em et al.},
who found a {\em unique} energy for which the same perfect transmission
took place in the RDM.

We now proceed to the problem of the disordered lattice, containing a
certain number of pair defects. To this end, we go back to Eq.\
(\ref{Sch}) and introduce the reflection and transmission amplitudes
through the relationships:
\begin{equation}
\label{amps}
\Psi(x) = \left\{ \begin{array}{ll} e^{iqx}+R_N\>e^{-iqx}, &
\mbox{if $x<1$,} \\ T_N\>e^{iqx}, & \mbox{if $x>N$,} \end{array}
\right.
\end{equation}
where $T_N$ and $R_N$ are the transmission and the
reflection amplitudes of a system with $N$ scatterers respectively,
$q\equiv\sqrt{E}$, and we
have put the lattice spacing $a=1$ without loss of generality. It is not
difficult to compute recursively both amplitudes using well-known
transfer matrix procedures (see, e.g., \cite{Kirkman}). In particular,
the transmission amplitude can be written as
\begin{equation}
\label{trans}
A_N=\left(\alpha_N+{\alpha^*_{N-1}\beta_N\over\beta_{N-1}}\right)\>
A_{N-1} - \left({\beta_N\over\beta_{N-1}}\right)\>A_{N-2},
\end{equation}
where $A_N\equiv 1/T^*_N$, and
\begin{equation}
\label{alphabeta}
\alpha_j\equiv\left[1-i\left({1\over 2q}\right)\>\lambda_j\right]
e^{iq}, \mbox{\ }
\beta_j\equiv -i\left({1\over 2q}\right)\>\lambda_j
e^{-iq}.
\end{equation}
Finally, Eq.\ (\ref{trans}) must be supplemented by the initial
condition $A_0=1$, $A_1=\alpha_1$ to completely determine the
amplitudes.

Once we have computed the transmission amplitude, some physically
relevant magnitudes can be readily obtained from it. Thus, the
transmission coefficient is given by
\begin{equation}
\label{transcoef}
\tau_N=|T_N|^2,
\end{equation}
whereas the resistivity, according to the Landauer formula
\cite{Landauer}, is simply
\begin{equation}
\label{res}
\rho_N={1\over |T_N|^2} -1.
\end{equation}
Aside from these two quantities, there are others that can also be
obtained from the transmission amplitude, although somewhat less
naturally. Indeed, the Lyapunov coefficient (which is nothing but
the inverse of the localization length) depends on this amplitude
through the expression \cite{Kirkman}
\begin{equation}
\label{Lyap}
\gamma_N={1\over 2N}  \log |T_N|^2  =
- {1\over 2N}  \log \tau_N,
\end{equation}
and it can be also shown \cite{Kirkman} that
the integrated density of states (IDOS) is related to $T_N$ by
\begin{equation}
\label{IDOS}
\Gamma_N=-{i\over 2\pi N} \log {T^*_N\over T_N};
\end{equation}
from this last expression, the density of states (DOS) can be obtained
by simple derivation with respect to the energy.

The results we have obtained so far provide an exact, although non
closed analytical description of any one-dimensional, disordered KP
model.
With them, we can compute the magnitudes we mentioned above for any
given model and, in particular, for the CRDM. All expressions are very
simple and suitable for an efficient numerical treatment of any specific
case. We will now evaluate them for several of these cases to check
whether there is any relevant feature of the transmission coefficient
and related quantities that may be the fingerprint of extended states.
We have to notice that there are several parameters that can be varied
in the CRDM: the strengths of the two kinds of scatterers, $\lambda$ and
$\lambda'$, the defect concentration, and the length of the system, $N$.
As to the first two of them, it can be checked that the factor $\lambda$
can be rescaled and subsequently suppressed in Eq.\ (\ref{Sch}), and
therefore, the relevant quantity is just the ratio $\lambda'/\lambda$,
which allows us to fix $\lambda=1$ from now on.

We first describe our results realizationwise, because we believe that
these are the most physically relevant; we briefly deal with the average
properties of the model below. In Fig.\
\ref{fig1}, we show the transmission coefficient for a system with
$\lambda'=1.5$, 5~000 scatterers, and a probability of having a dimer
$q=0.5$ \cite{nota}. In this plot, it is clearly appreciated the peak in
the
transmission coefficient very close to the predicted value for the first
resonance ($E\approx 3.7626$ for this parameter set). Moreover,
neighbouring states have a transmission coefficient close too unity,
that decreases as we move away from the resonance. In Fig.\
\ref{fig2}, the
Lyapunov exponent is plotted vs the energy for the same system; again,
we appreciate that there is a deep minimum around the resonance value,
which
implies a very large localization length, much larger than the system
size.
The other magnitude we study, the resistivity, confirms the existence of
an energy interval for which a typical realization of our model shows
transport properties similar to those of perfect lattices.

The IDOS, which is plotted in Fig.\ \ref{fig3}, deserves some separate
comments. Due to the presence
of the multivalued $\arctan$ function in the defining relationship
(\ref{IDOS}), this magnitude is very sensitive to the resolution in
energies: if there is a jump in the $\arctan$ between two points for
which the IDOS is computed, this jump will be missed and the IDOS will
be subsequently underestimated. However, we checked several cases
computing the IDOS with tiny
energy steps ($5\times 10^{-6}$) which is very time consuming; with this
accuracy, we recover the agreement between systems of different sizes
(notice that the magnitude we discuss is in fact the IDOS per volume) as
regards the total number of states and the IDOS structure. As to this
last feature, we want to stress that the IDOS is well behaved (smooth)
over all the studied range of energies. This implies that the same
argument used by Dunlap {\em et al.}\ \cite{Dunlap} to show that
$\sqrt{N}$ states were extended holds in this case too, because the
reasoning depends crucially on the DOS structure (see \cite{Bovier}).

It is most important to report on how the above picture is modified when
the system parameters are changed. First of all, the main characteristic
of our model, i.e., that it has an infinite number of resonances, is
confirmed by our calculations; besides, the higher the resonant energy
(meaning the higher $n$ in $[n\>\pi/2,(n+1)\>\pi/2]$), the wider the
peak in the transmission coefficient and the other transport magnitudes.
The peak width increases also when
decreasing $\lambda'$ towards $\lambda=1$, and decreases when increasing
$\lambda'$ up to its maximumn $\lambda'=2$. This is to be expected,
because when $\lambda'=\lambda$ we recover the perfect lattice. With
respect to the other parameters, the number of scatterers and the
concentration of dimers, both cause a narrowing of the set of extended
states when they are increased in the studied range ($100\leq
N\leq 50~000$, $0.1\leq q\leq 0.5$), although it is important to stress
that
this set always has nonzero width. Interestingly, when the number of
scatterers increases, the IDOS steepens (i.e., the DOS exhibits a
sensitive increment) around
the resonant energy; consequently, the number of extended states may be
constant in spite of the decreasing of the width of the transmission
peak.

We now comment on the averaged results.
When computing averages, they were taken over a number of
realizations varying from 100 to 10 000 to check the convergence of the
computed mean values. The convergence was always satisfactory, with
discrepancies of less than 1\% between all the ensembles. Once more,
however, to get accurate results for the IDOS is quite time consuming
due to the necessary resolution in energies. The averaged results for
the transmission coefficient, the Lyapunov
exponent, the resistivity, and the density of states are basically the
same as those for a typical realization we commented on above. This is
a crucial point because it supports our claim that those are the main
features
of our model irrespective of the particular realization of the disorder.

In summary, we have studied a Kronig-Penney model with two kinds of
$\delta$-functions, one of them constrained to appear only pairwise.
This
is a continuous extension of the RDM of Dunlap {\em et al.}
\cite{Dunlap}.
We find an infinite number of energies for which the reflection
coefficient of a single defect vanishes. We have shown, through
numerical evaluation of exact expressions, that these resonances give
rise to a
very large number of extended states, that can be much larger than that
of the RDM where a unique resonance exists. These extended states are
characterized by a transmission coefficient close to unity and a
localization
length much larger than the system one. The basis for the existence
of a extended states as relevant as to affect the transport properties,
the smooth character of the DOS around the resonance
\cite{Dunlap,Bovier}, holds, supporting our conclusions. The increasing
of the DOS around the resonant for large systems helps keep
relevant the number of extended states. We believe that similar results
will arise in related continuous models. Work currently in progress
regarding
the structure of the wavefunctions, as well as the development of
better, more accurate methods to compute the DOS, will be reported
elsewhere.

\bigskip


We are much deeply indebted to Sergey A.\ Gredeskul for very helpful
discussions on our calculations, as well as to Rainer Scharf for drawing
our attention on the Random Dimer model. All computations have been
carried out using facilities of the Servicios
Inform\'aticos of the Universidad Carlos III de Madrid.
A.\ S.\ was partially supported by C.I.C.\
y T.\ (Spain) project MAT90-0544.


\begin{figure}
\caption{Transmission coefficient vs energy for a system with
$\lambda'=1.5$, 5 000 scatterers, and a probability of having a dimer of
0.5 \protect\cite{nota}. The arrow marks the predicted resonance.}
\label{fig1}
\end{figure}
\begin{figure}
\caption{Lyapunov exponent vs energy for a system with
$\lambda'=1.5$, 5 000 scatterers, and a probability of having a dimer of
0.5 \protect\cite{nota}.
The dashed line marks the inverse of the
system length: Energies with a lower exponent will have a localization
length larger than the system one.
The arrow marks the predicted resonance.}
\label{fig2}
\end{figure}
\begin{figure}
\caption{Integrated density of states for systems with
$\lambda'=1.5$ and a probability of having a dimer of
0.5 \protect\cite{nota}, of sizes 1 000, 5 000, and 10 000 scatterers
from top to bottom. The arrow marks the predicted resonance.}
\label{fig3}
\end{figure}


\begin{references}
\bibitem{nota} The probability of having a dimer is used when building
up the system; each site is assigned a value $\lambda$ or $\lambda'$
according to it. It does not coincide with the concentration, however,
because if a site happens to be part of a dimer, the next one is
automatically part of a dimer too, so the value of the delta function at
that point is deterministic.
\end{references}
\end{document}